\documentclass[lettersize,journal]{IEEEtran}
\usepackage{amsmath,amsfonts}
\usepackage{algorithmic}
\usepackage{algorithm}
\usepackage{array}
\usepackage{textcomp}
\usepackage{stfloats}
\usepackage{url}
\usepackage{verbatim}
\usepackage{graphicx}
\usepackage{cite}
\usepackage{subfigure}
\begin{document}

\title{The Simons Observatory: Characterization of All DC/RF Routing Wafers for Detector Modules}

\author{Alicia Middleton, Kyuyoung Bae, Cody J. Duell, Shannon M. Duff, Erin Healy, Zachary B. Huber, Johannes Hubmayr, Ben Keller, Lawrence T. Lin, Michael J. Link, Tammy J. Lucas,  Michael D. Niemack, Eve M. Vavagiakis, Yuhan Wang

\thanks{A. Middleton, C. J. Duell, Z. B. Huber, B. Keller, L. T. Lin, M. D. Niemack, and Y. Wang are with the Department of Physics, Cornell University, Ithaca, NY 14853, USA. (e-mail: amm542@cornell.edu).

K. Bae is with the Quantum Sensor Division, National Institute of Standards and Technology, Boulder, CO 80305, USA, and the Department of Physics, University of Colorado, Boulder, CO 80309, USA.

S. M. Duff, J. Hubmayr, M. J. Link, and T. J. Lucas are with the Quantum Sensor Division, National Institute of Standards and Technology, Boulder, CO 80305, USA.

E. Healy is with the Kavli Institute of Cosmological Physics, University of Chicago, Chicago, IL, 60607, USA.

E. Vavagiakis is with the Department of Physics, Duke University, Durham, NC, 27704, USA, and the Department of Physics, Cornell University, Ithaca, NY 14853, USA.
}}



\maketitle

\begin{abstract}
The Simons Observatory (SO) is a cosmic microwave background experiment with over 67,000 polarization-sensitive transition-edge sensor (TES) detectors currently installed for use in observations and plans to increase the total detector count to $\sim$98,000 detectors with the Advanced SO upgrade.  The TES arrays are packaged into Universal Focal-Plane Modules (UFMs), which also contain the multiplexing readout circuit.  Within a readout module, a DC/RF routing wafer provides a cold interface between the detectors and the readout multiplexing chips. Each routing wafer hosts twelve bias lines, which contain the $\sim$400\,$\mu \Omega$ shunt resistors that are part of the TES bias circuitry. More than 70 routing wafers have been fabricated and tested both at room temperature and 100 mK before integration into UFMs.  The lab measurements for all screened wafers have been compiled to show the distribution of measured average shunt resistance $R_{sh}$ for each bias line, both across bias lines on a single routing wafer and across all routing wafers.  The mean average shunt resistance for all wafers was found to be 396\,$\mu \Omega$ with a standard deviation of 16\,$\mu \Omega$, or $\sim$$4 \%$.  For each wafer, we note good uniformity of average $R_{sh}$ between bias lines, with a slight downward trend with increasing distance from the center of the wafer.  The fabrication data collected at room temperature shows agreement with the cryogenic measurements of $R_{sh}$ distribution.
\end{abstract}

\section{Introduction}

The Simons Observatory (SO) is a cosmic microwave background (CMB) experiment located in the Atacama Desert in Chile at an elevation of 5200\,m.  In its current configuration, SO consists of three 0.5\,m Small-Aperture Telescopes (SATs) and one 6\,m  Large-Aperture Telescope (LAT) with a total of $\sim$67,000 transition-edge sensor (TES) detectors between all receivers installed for astronomical observations at wavelengths spanning the range 27$\textendash$280\,GHz.  The receivers for each of the SATs have one optics tube, while the LAT receiver (LATR) has six, but has space for up to 13.  The LATR will be fully populated with optics tubes as part of the Advanced SO (ASO) upgrade, which will increase the number of on-sky detectors to $\sim$98,000.  The wide range of science goals for SO include constraining cosmological parameters and neutrino mass, surveying galaxy clusters out to $z > 2$, and providing samples of active galactic nuclei, dusty star-forming galaxies, and transient astronomical signals \cite{Ade_2019}\cite{thesimonsobservatorycollaboration2025simonsobservatorysciencegoals}.

SO observes in six frequency bands with three distinct dichroic TES array designs.  Low-Frequency (LF) arrays have bands centered at approximately 30 and 40\,GHz, Mid-Frequency (MF) arrays at 90 and 150\,GHz, and Ultra-High-Frequency (UHF) arrays at 220 and 280\,GHz.  Each MF and UHF array has 430 optically-coupled pixels with four TES detectors per pixel as well as 36 dark TES detectors for calibration, giving a total of 1,756 detectors per wafer.  A full description of fabrication details of the MF and UHF arrays can be found in \cite{Duff_2024}.  The detector arrays are read out using microwave superconducting quantum interference device (SQUID) multiplexing \cite{10.1063/5.0033416} \cite{Jones_2024}.

\begin{figure}
    \includegraphics[width= 3.5in]{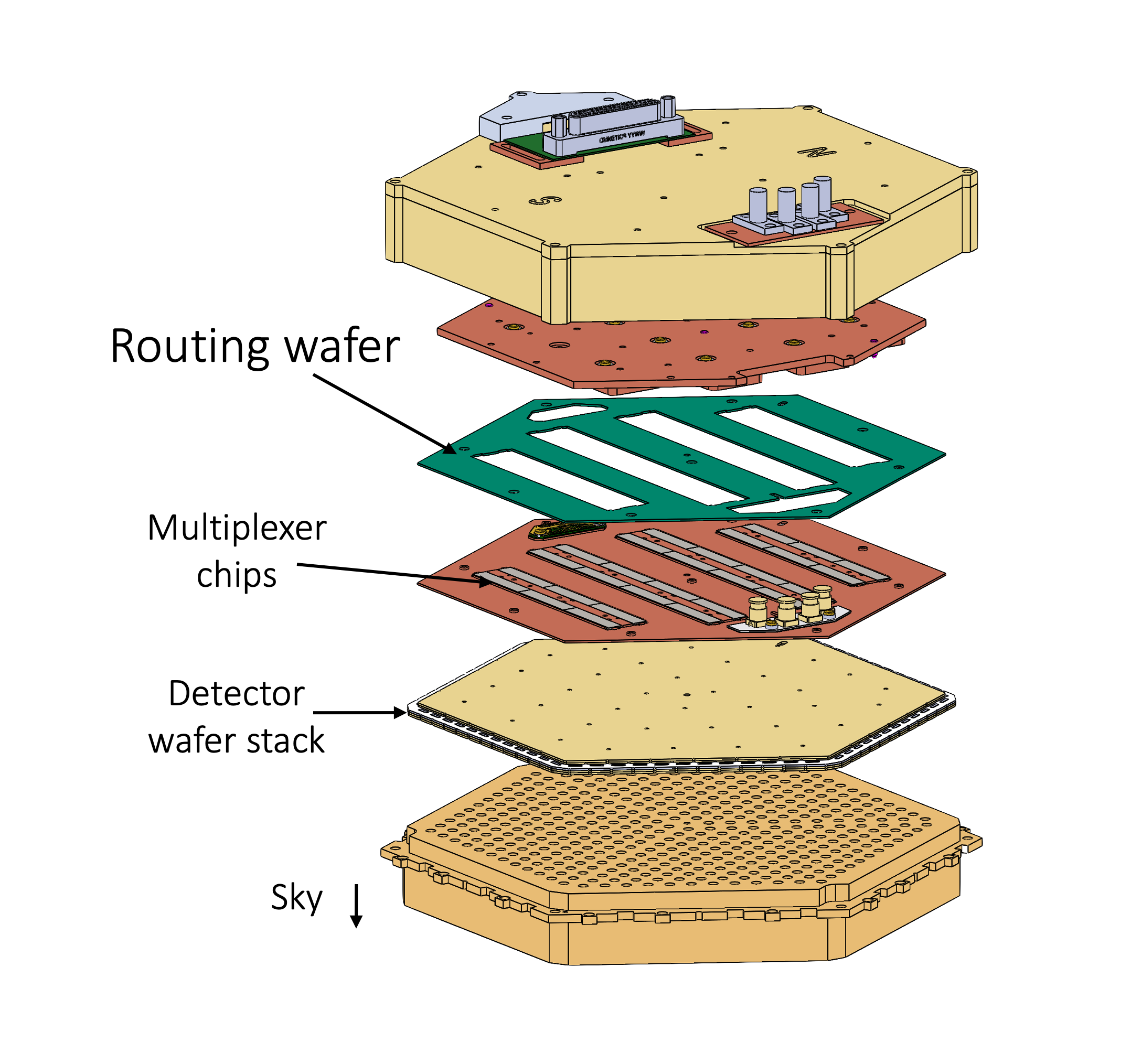}
    \caption{Exploded view of an SO UFM showing the optical coupling and multiplexing readout components\cite{mccarrick202190150ghzuniversal} \cite{Healy_2022} including the location of the routing wafer, shown in green.  The cutouts in the routing wafer design leave open space for the multiplexer chips as well as PCBs and dowel pins for mechanical assembly (see Figure \ref{fig:waferdiagram}).}
    \label{fig:UFM}
\end{figure}

The TES arrays for SO are packaged into Universal Focal Plane Modules (UFMs) \cite{mccarrick202190150ghzuniversal} \cite{Healy_2022}, which also contain the optical coupling and multiplexing cold readout \cite{McCarrick_2021}.  An exploded diagram of the components of a UFM is shown in Figure \ref{fig:UFM}.  The focal plane of each SAT consists of seven tiled UFMs, while the focal plane of each optics tube in the LATR consists of three UFMs.  A DC/RF routing wafer, colored green in Figure \ref{fig:UFM}, serves as the cold interface between the detector array and readout multiplexing chips in a UFM. 

\begin{figure*}
\label{fig:circuitry}
\includegraphics[width = \textwidth]{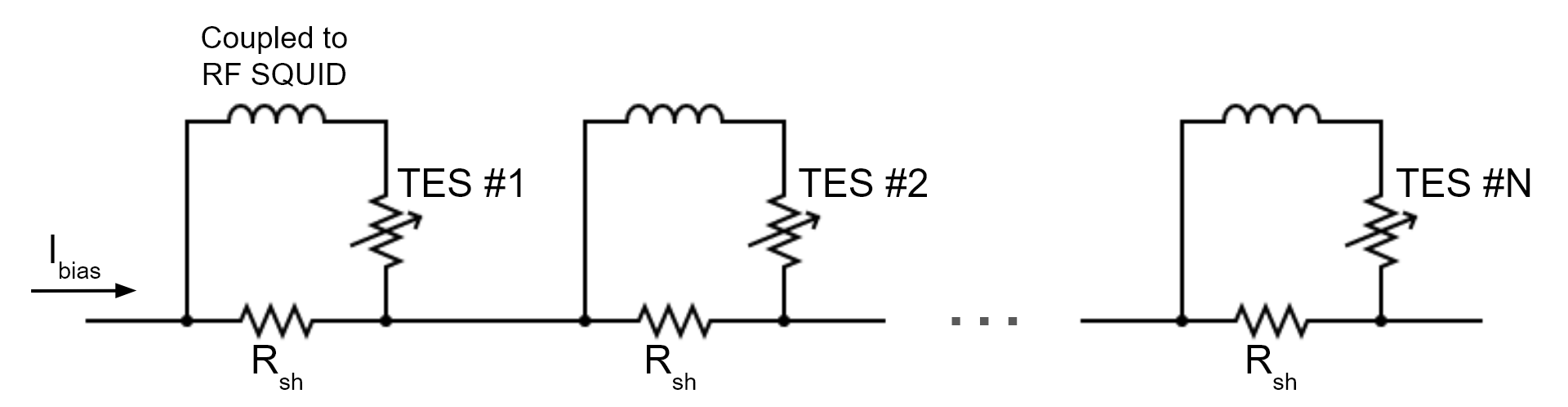}
\caption{The TES bias circuitry and routing wafer bias line shunt resistors.  Each TES is in parallel with a low-impedance shunt resistor so that the bias voltage can be set with a current, $I_{bias}$, flowing into the circuit and in series with an inductor which couples to an RF-SQUID so that the current of the TES is read out.}
\end{figure*}

\begin{table}[!t]
\caption{Number of Channels per Bias Line\label{tab:table1}}
\label{tab:bl_channels}
\centering
\def\arraystretch{1.5}%
\begin{tabular}{|c|c|}
\hline
Bias Line & channels \\
\hline
0 & 128 \\
\hline
1 & 172 \\
\hline
2 & 144 \\
\hline
3 & 144 \\
\hline
4 & 172 \\
\hline
5 & 128 \\
\hline
6 & 128 \\
\hline
7 & 160 \\
\hline
8 & 150 \\
\hline
9 & 150 \\
\hline
10 & 160 \\
\hline
11 & 128 \\
\hline
total & 1764 \\
\hline

\end{tabular}
\end{table}

The TES detectors are voltage-biased through a low-impedance shunt resistor, which sets the voltage via a bias current supplied to the TES bias circuit as shown in Figure 2. Each routing wafer hosts twelve bias lines which contain these shunt resistors, with a 400\,$\mu\Omega$ shunt resistance chosen to be $\sim$$10\%$ of the TES operating resistance \cite{McCarrick_2021}.  It is important that this value is consistent across all bias lines on all routing wafers, since meeting nominal shunt resistance enables the noise equivalent power (NEP) requirement to be met for the SO TES detectors and uniformity will ensure reliability of the detectors.  The number of channels that are read out by each bias line is shown in Table \ref{tab:bl_channels}.  We note that each MF and UHF 
TES array has 1,756 detectors, with two pixels (and therefore space for eight detectors) being used for wafer-aligning dowel pins. 

\begin{figure}[b!] 
    \includegraphics[width= 3.5in]{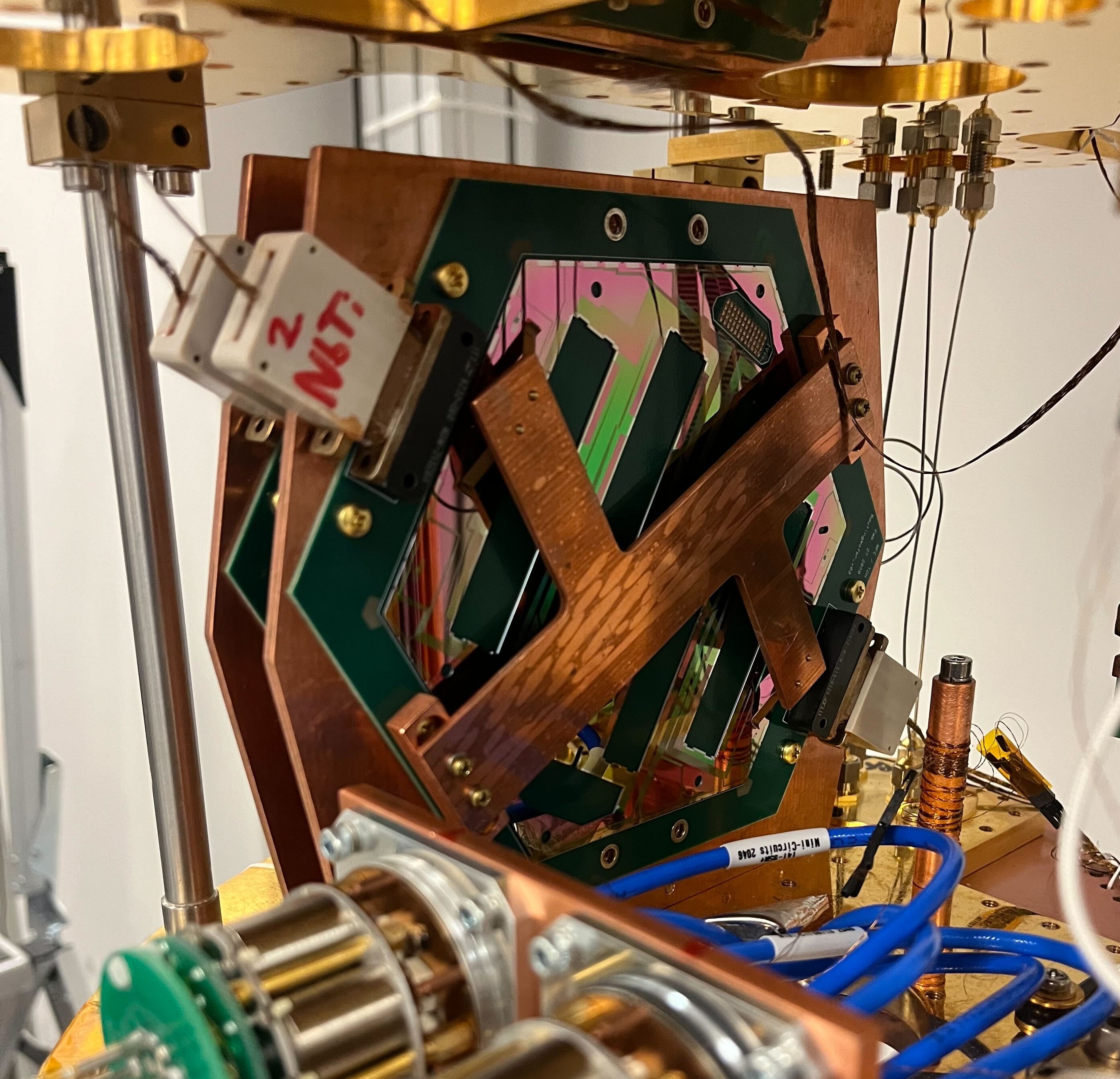}
    \caption{A routing wafer mounted onto a PCB and installed in a Bluefors LD400 dilution refrigerator for cryogenic testing.}
    \label{fig:mountedwafer}
\end{figure}

In the following section, we describe the cryogenic cold screening of each routing wafer fabricated for SO required before integration into UFMs.  This cryogenic testing includes checking for shorts between bias lines and measurements of the average shunt resistance, $R_{sh}$, of each bias line.  Then, we discuss the distribution of $R_{sh}$ across all wafers, and compare the results to room temperature fabrication metrology data.  Finally, we conclude and summarize the results of our routing wafer screening measurements.

\section{Experimental Design}

More than 70 routing wafers have been fabricated on 150\,mm silicon wafers at NIST in Boulder, CO and tested both at room temperature and cryogenic temperatures of 100\,mK before integration into UFMs.

For cryogenic cold screening measurements, the routing wafers were mounted onto printed circuit boards (PCBs) and installed in a Bluefors LD400 dilution refrigerator, as shown in Figure \ref{fig:mountedwafer}, to be cooled to the UFM operating temperature of $\sim$100\,mK.  To test critical current, we found the total resistance of each bias line using a four-lead measurement with a LakeShore 370 at five current values ranging 0.0316$\textendash$10\,mA.  We then divided the total bias line resistance by the number of shunt resistors in that bias line, as shown in Table \ref{tab:bl_channels}, to find the average $R_{sh}$ at each current value, and these values are averaged to find the overall average $R_{sh}$ for each bias line.  All bias lines were also checked for shorts at both room temperature and at 100\,mK.  Other room temperature measurements performed during fabrication at NIST are described in Section 3C.

Routing wafers are excluded from integration into UFMs if bias line yield is $< 90\%$ (i.e. two or more bias lines are unusable due to shorts or going normal during critical current testing).  Of tested routing wafers, $\sim$$18\%$ of wafers had at least one short and $\sim$$4\%$ of all bias lines were found to be shorted to at least one other bias line.  $\sim$$18\%$ of wafers were shown to have at least one bias line go normal, with normal bias lines accounting for $\sim$$2\%$ of all measured bias lines.  Bias lines that went normal during cryogenic testing, meaning that they had low critical current and therefore very high $R_{sh}$ values, are not included in the $R_{sh}$ statistics described in the following section.  These bias lines are excluded from detector operations.  

\section{Results and Analysis}

\subsection{$R_{sh}$ Distribution}

\begin{figure}[b!] 
    \includegraphics[width= 3.5in]{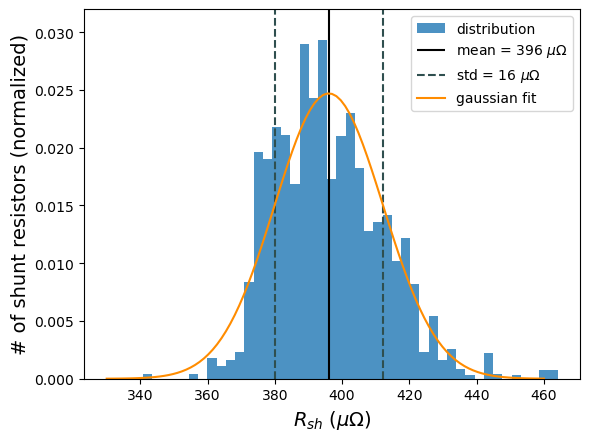}
    \caption{The distribution of all measured average bias line shunt resistances.  Each average $R_{sh}$ measurement is weighted by the number of shunt resistors in the bias line (see Table \ref{tab:bl_channels}), and the y-axis is normalized such that the area under the curve is one.  The distribution has a mean value of 396\,$\mu \Omega$ and with a standard deviation of 16\,$\mu \Omega$, or $\sim$$4 \%$.  This is within agreement of the nominal 400\,$\mu \Omega$ $R_{sh}$ for SO TES detectors.}
    \label{fig:Rshuntdistribution}
\end{figure}

The distribution of all measured average bias line shunt resistances on all wafers is shown in Figure \ref{fig:Rshuntdistribution}.  The measured $R_{sh}$ value for each bias line is weighted by the number of shunt resistors on the bias line, as given in Table \ref{tab:bl_channels}. The mean $R_{sh}$ value was found to be 396 $\mu \Omega$ with one standard deviation of 16\,$\mu \Omega$, or $\sim$$4 \%$.  This is in agreement with the nominal $R_{sh}$ value of 400\,$\mu \Omega$ that is used in NEP calculations for the SO TES detectors.  

\subsection{Radial Dependence of $R_{sh}$}

\begin{figure}[t!] 
    \includegraphics[width= 3.5in]{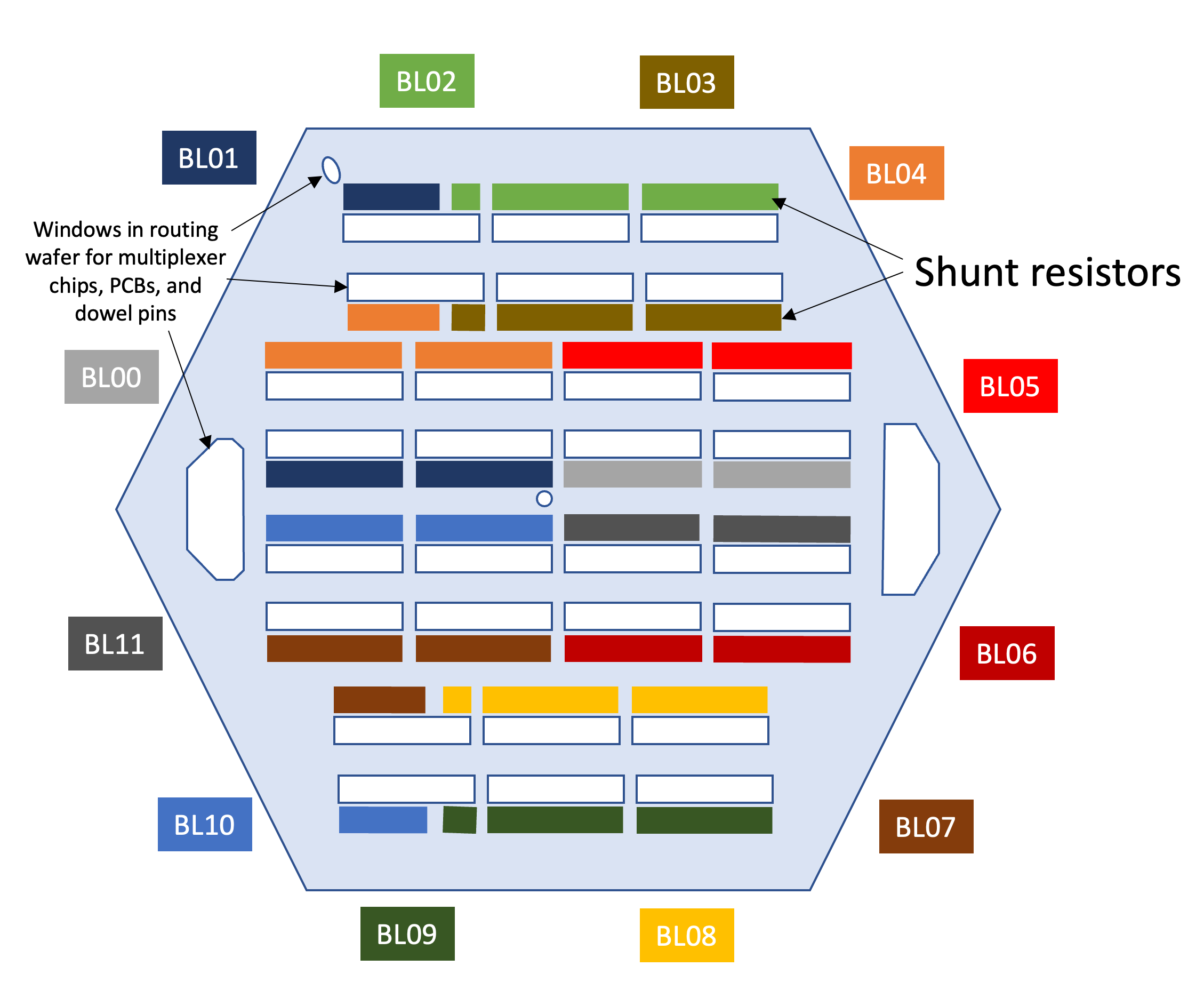}
    \caption{Routing wafer diagram showing approximate locations of shunt resistors for each bias line, highlighted in color by bias line.  Bias lines were divided into pairs based on relative distance from the center of the wafer so that we could analyze radial dependence of $R_{sh}$.}
    \label{fig:waferdiagram}
\end{figure}

A diagram of a routing wafer showing the approximate locations of the shunt resistors for each bias line is shown in Figure \ref{fig:waferdiagram}.  We divided bias lines into pairs based on the relative distances of their shunt resistors from the center of the wafer, with smallest to largest radius as follows:

\begin{itemize}
  \centering
  
  \item (BL00, BL11)
  \item (BL01, BL10)
  \item (BL05, BL06)
  \item (BL04, BL07)
  \item (BL03, BL08)
  \item (BL02, BL09)
  
\end{itemize}

\noindent We then found the on-wafer average distance of the shunt resistors from the center of the wafer and extracted the $R_{sh}$ distribution across all wafers for each bias line pair.  

\begin{figure}[t!]
    \centering
    \begin{subfigure}
        \centering
        \includegraphics[width=3.5in]{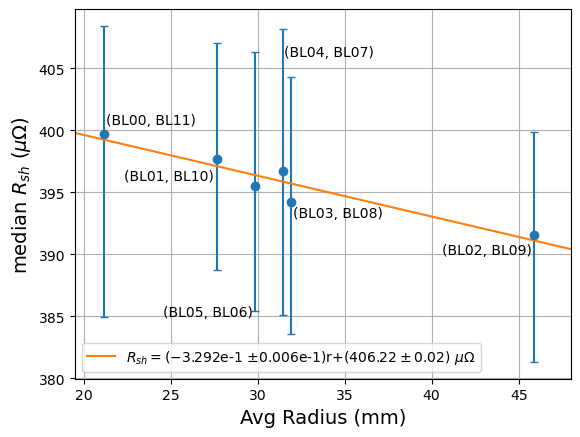}
    \end{subfigure}%

    \begin{subfigure}
        \centering
        \includegraphics[width=3.5in]{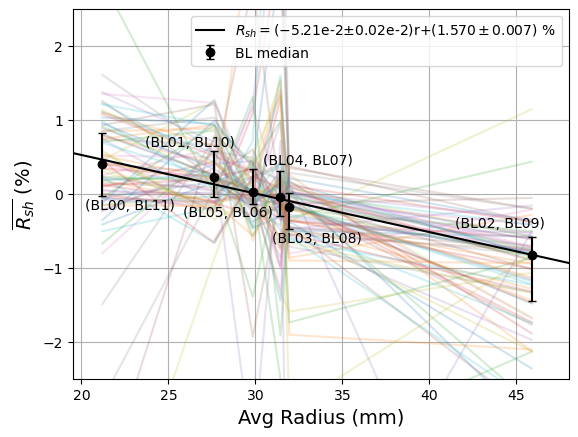}
    \end{subfigure}
    \caption{Top: Median $R_{sh}$ for each bias line pair plotted against the average radius of shunt resistors in that bias line pair.  Error bars represent the 25th and 75th percentiles of the distribution for each bias line pair.  A slight negative correlation of $R_{sh}$ with radius is apparent, but this variation is smaller than the $R_{sh}$ variation between all wafers. Bottom: The fractional variation $\overline{R_{sh}}$ of the measured $R_{sh}$ of each bias line pair with respect to the median $R_{sh}$ for the corresponding wafer.  Each colored line represents the $R_{sh}$ measurements for one wafer, and the solid black points give the median value for the distribution at that bias line pair.  Error bars again show the 25th and 75th percentiles of the distribution.  This plot further highlights the correlation of $R_{sh}$ with distance from the center of the wafer, but again, this variation is smaller than the variation between individual wafers.}
    \label{fig:Rshuntradius}
\end{figure}

The distribution of $R_{sh}$ values for each bias line pair plotted against the average distance of shunt resistors in that pair from the center of the wafer is shown in the top panel of Figure \ref{fig:Rshuntradius}, where error bars represent the 25th and 75th percentiles.  A negative correlation of $R_{sh}$ with radius is apparent.

The bottom panel of Figure \ref{fig:Rshuntradius} shows the variation of $R_{sh}$ across each individual wafer.  The y-axis gives the fractional variation of the average of each bias line pair from the median $R_{sh}$ value for that wafer, such that

\begin{equation}
    \overline{R_{sh}} = \frac{R_{sh, BL} - \text{median }R_{sh}}{\text{median }R_{sh}} \times 100
\label{eq:fractionalrshunt}
\end{equation}

\noindent for each bias line pair, resulting in a distribution about zero for each wafer.  The colored lines represent the $R_{sh}$ measurements for individual wafers, while the solid black points give the median value of the distribution for each bias line pair, again using the 25th and 75th percentiles as error bars.  This plot further highlights the negative correlation of $R_{sh}$ with distance from center as seen within individual wafers.  We see a typical variation of $< 2\%$ across one wafer due to this negative correlation with radius, which is smaller than the $\sim$$4\%$ variation in $R_{sh}$ between wafers (see Figure \ref{fig:Rshuntdistribution}).

\subsection{Comparison to Room Temperature Data}

\begin{figure}[t!]
    \centering
    \begin{subfigure}
        \centering
        \includegraphics[width=3.5in]{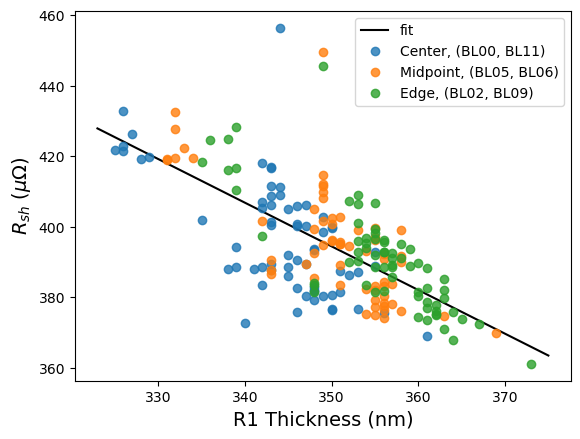}
    \end{subfigure}%

    \begin{subfigure}
        \centering
        \includegraphics[width=3.5in]{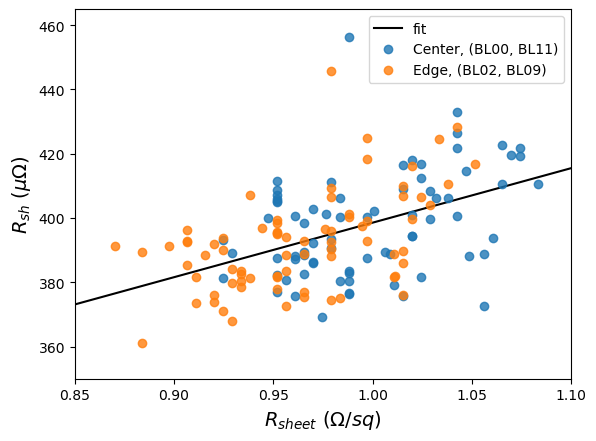}
    \end{subfigure}
    \caption{Top: Comparison of the average $R_{sh}$ of the labeled bias line pair on each wafer with profilometry measurements of the R1 (shunt resistor) layer of that wafer. Each point represents the average value of $R_{sh}$ for the corresponding bias line pair at the center, midpoint, and edge for one wafer. As expected, we observe a negative correlation of $R_{sh}$ with R1 layer thickness.  The best fit line, in black, is given by $R_{sh} = (-1.23878 \pm 0.00008)R1 + (828.08 \pm 0.03)$\,$\mu \Omega$. Bottom: Comparison of $R_{sh}$ with $R_{sheet}$ measurements at room temperature.  Each point again gives the measurements for the bias line pair at the center and the edge of the wafer.  As expected, we observe a positive correlation of $R_{sh}$ measured at cryogenic temperatures with $R_{sheet}$ measured at room temperature. The best fit line, in black, is given by $R_{sh} = (169.51 \pm 0.02)R_{sheet} + (229.04 \pm 0.02)$\,$\mu \Omega$.} 
    \label{fig:roomtempdata}
\end{figure}

We focus on two room temperature measurements performed during fabrication at NIST: (1) profilometry of the shunt resistor PdAu layer, designated as R1, at the center, edge, and midpoint of the wafer, and (2) measurements of sheet resistance ($R_{sheet}$) of the R1 layer at the center and edge of the wafer.  $R_{sheet}$ is extracted from a four-wire electrical measurement using a cross-bridge resistor \cite{Walton1999MICROELECTRONICTS}.  For comparison of $R_{sh}$ to room temperature measurements, we selected the bias line pairs with shunt resistors located at approximately the center, midpoint, and edge of a wafer.

A comparison to fabrication data measured at room temperature is shown in Figure \ref{fig:roomtempdata}.  The top panel compares the average measured $R_{sh}$ values for each wafer to the R1 layer thickness at the center, midpoint, and edge.  Bias line pairs were selected which approximately correspond to these locations and the average value of the pair at each location was plotted, with each point giving the value for one wafer.  The bottom panel compares the average $R_{sh}$ values for the bias line pairs at the center and edge to the measured $R_{sheet}$ values at those locations.  Each point again represents the measurements for one wafer.  As expected, we observe a negative correlation of $R_{sh}$ with R1 thickness, and a positive correlation of $R_{sh}$ with $R_{sheet}$.  Therefore, room temperature fabrication data is shown to be in agreement with cryogenic cold screening measurements.

\section{Conclusions}

We measured good uniformity in $R_{sh}$ across all routing wafers and bias lines.  The measured average $R_{sh}$ for all bias lines is 396\,$\mu \Omega$ with a standard deviation of 16\,$\mu \Omega$ or $\sim$$4 \%$, which is within excellent agreement of the nominal $\sim$400\,$\mu \Omega$ $R_{sh}$ that enables the NEP requirement to be met for the SO TES arrays.  A negative correlation of $R_{sh}$ with the distance of the shunt resistors from the center of each wafer is apparent, but this variation is smaller than the variation between individual wafers.  Cryogenic measurements of $R_{sh}$ also show agreement with the room temperature fabrication measurements of R1 layer thickness and $R_{sheet}$ at corresponding locations on the wafer.  Following warm and cold screening, the routing wafers are integrated into UFMs as part of the multiplexing cold readout for the SO TES arrays.  All fabrication and screening of routing wafers has been completed for SO and ASO.

\section{Acknowledgments}

This work was supported in part by a grant from the Simons Foundation (Award 457687, B.K.). This work was supported by the U.S. National Science Foundation (Award Number: 2153201).

\bibliographystyle{IEEEtran}

\nocite{*}

\bibliography{bibliography}


\vfill

\end{document}